\documentclass[journal]{IEEEtran}
\usepackage{graphicx}
\usepackage[caption=false, font=footnotesize]{subfig}
\usepackage{cite}
\usepackage{amssymb, amsmath}
\usepackage{multirow}
\usepackage{color}

\begin{document}

\title{Task-Oriented Wireless Communications for Collaborative Perception in \\Intelligent Unmanned Systems}
% \title{Collaborative Perception Oriented Wireless Communications for Intelligent Unmanned Systems}

%\title{Collaborative Perception for Intelligent Unmanned Systems with Wireless-in-the-Loop}

\author{Sheng~Zhou,~\IEEEmembership{Senior Member,~IEEE,} Yukuan Jia, Ruiqing Mao, Zhaojun Nan,~\IEEEmembership{Member,~IEEE,}\\ Yuxuan~Sun,~\IEEEmembership{Member,~IEEE,} Zhisheng~Niu,~\IEEEmembership{Fellow,~IEEE}
\thanks{Sheng~Zhou, Yukuan Jia, Ruiqing Mao, Zhaojun Nan and Zhisheng~Niu are with Beijing National Research Center for Information Science and Technology, Department of Electronic Engineering, Tsinghua University, Beijing 100084, China.}
\thanks{Yuxuan~Sun (\emph{Corresponding Author}) is with School of Electronic and Information Engineering, Beijing Jiaotong University, Beijing 100044, China.}
%\thanks{This work is sponsored in part by ...}
}

\maketitle

\begin{abstract}
Collaborative Perception (CP) has shown great potential to achieve more holistic and reliable environmental perception in intelligent unmanned systems (IUSs). However, implementing CP still faces key challenges due to the characteristics of the CP task and the dynamics of wireless channels. In this article, a task-oriented wireless communication framework is proposed to jointly optimize the communication scheme and the CP procedure. We first propose channel-adaptive compression and robust fusion approaches to extract and exploit the most valuable semantic information under wireless communication constraints. We then propose a task-oriented distributed scheduling algorithm to identify the best collaborators for CP under dynamic environments. The main idea is learning while scheduling, where the collaboration utility is effectively learned with low computation and communication overhead. Case studies are carried out in connected autonomous driving scenarios to verify the proposed framework. Finally, we identify several future research directions.
\end{abstract}

\section{Introduction}

Intelligent unmanned systems (IUSs) have received widespread attention in recent years, and play increasingly significant roles in various aspects of human life and society, such as transportation, surveillance, and industry. In IUSs, agents such as autonomous vehicles, unmanned aerial vehicles (UAVs) and robots are typically equipped with a variety of sensors, including cameras, LiDARs, and millimeter-wave radars. They need to carry out perception tasks such as object detection, tracking, and semantic segmentation using multi-modal sensor data and advanced artificial intelligence (AI) algorithms. Conventional stand-alone perception is susceptible to occlusions and ambiguity in complex scenarios, due to single viewpoint and limited sensing range. In \emph{collaborative perception} (CP), on the other hand, sensor data from multiple agents are fused to improve the perception quality of distant or occluded views \cite{Yang2021Machine}. For example, in object detection tasks, more true positive objects and fewer false positives are expected to be detected through CP \cite{where2comm}. For semantic segmentation tasks, CP is expected to improve the intersection-over-union between the ground truth and the predicted region of each object \cite{bevfusion}.

According to the data sharing and fusion stage, CP can be categorized into three levels: raw-level, feature-level, and object-level. 
In raw-level CP, agents directly exchange raw sensor data, such as images and point clouds. While these raw data preserve the full information for perception, they require extremely high communication costs. 
In object-level CP, each agent conducts perception tasks locally based on its own sensor data, and then exchanges the perception results, such as detected objects, with others. The communication cost is generally low, but many details in sensor data are lost. As a result, the perception quality is usually the lowest among the three categories.
In feature-level CP, agents extract features from their raw sensor data, exchange these features on-demand, and finally conduct fusion algorithms. 
Since feature-level CP can strike a balance between perception quality and communication cost, it is a promising solution for IUSs \cite{cp-ITSMag}.

Typically, agents in IUSs are moving, and thus they are connected wirelessly. 
%For example, autonomous vehicles drive at various speeds in complex urban areas, and a team of UAVs are deployed in aerial surveillance systems to detect and track objects. 
The wireless communication process plays a significant role in CP, aiming to deliver useful information among agents and to enhance the perception quality as much as possible. Different from conventional wireless communications, data sharing in CP needs to be perception-aware, taking into account the sensing and computing capabilities of agents as well as wireless channel qualities. Therefore, a novel \emph{task-oriented wireless communication framework} needs to be designed. In fact, CP is a representative example of the emerging paradigm in communication systems, namely \emph{task-oriented communications}~\cite{task-oriented} or \emph{semantic communications}~\cite{Kountouris2021semantics}, which has drawn great attention recently.

\begin{figure*}
	\centering
	\includegraphics[width=0.85\linewidth]{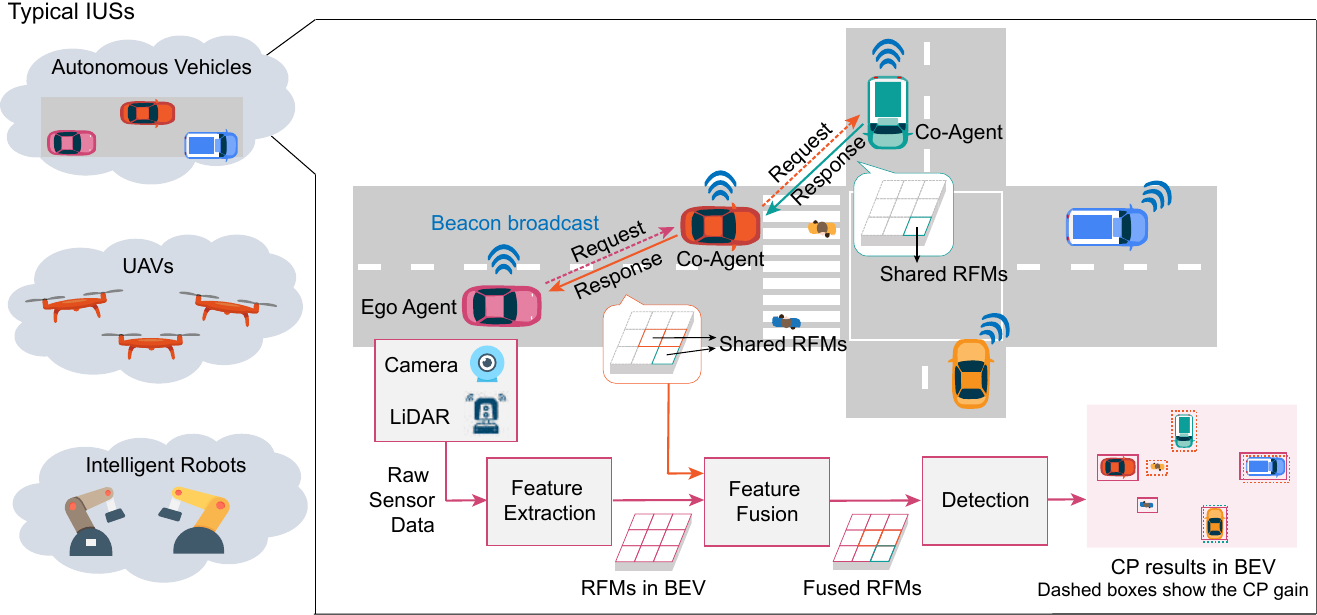}
	\caption{{Illustration of the proposed task-oriented wireless communication framework for CP in IUSs.}}
	\label{fig:cparch}
\end{figure*}

Despite the potential benefits, optimizing task-oriented wireless communications for CP is extremely challenging.
First, the \emph{wireless channel conditions} among the collaborative agents are dynamic, due to the time-varying path loss, shadowing, and fast fading.
% First, the communication topology of agents is dynamic, with time-varying channel conditions, intermittent connectivity, and limited bandwidth. 
This necessitates a channel-adaptive CP scheme. 
Second, the \emph{perception topology}, defined by whether an agent (or a group of agents) can monitor an area or detect an object of interest, is constantly changing. 
The mobility of agents causes dynamic perspectives of mounted sensors.
Therefore, it is challenging to determine which collaborative agents are more beneficial for enhancing the perception quality, taking the time-varying wireless channels into account.
%Multi-hop communication may also be employed when aggregating distant information.
Last but not least, features to be fused may not align in both temporal and spatial domains, due to the asynchrony in sensing, computing and communication delays, and localization errors. A robust feature fusion scheme needs to be incorporated.

In this article, we propose a task-oriented wireless communication framework, aiming at improving the CP quality of IUSs. The proposed framework adopts feature-level CP. First, given the collaborative agents and their wireless communication constraints, we design a channel-adaptive feature extraction module to extract lightweight regional feature maps (RFMs) from raw sensor data. Meanwhile, a deep metric learning-based robust fusion module is proposed, which matches the RFMs of different agents under inaccurate spatio-temporal alignment to improve the perception quality.
Then, in the theoretic framework of the restless combinatorial multi-armed bandit, we propose a task-oriented distributed scheduling algorithm to identify the best collaborators under the dynamic environment. The CP utility, which jointly reflects the perception and communication conditions, is effectively learned in an online fashion.
Through two case studies in connected autonomous driving scenarios, we show the potential perception gain of the proposed framework and stimulate further research directions.

\section{Task-Oriented Wireless Communication for CP: An Overview}

In this section, we propose a task-oriented wireless communication framework that jointly optimizes the communication scheme and the perception procedure for higher CP quality.

As shown in Fig. \ref{fig:cparch}, we adopt a \textit{pull-based distributed} communication framework. Each agent simultaneously acts as an ego agent and a collaborative agent (co-agent). The co-agents periodically beacon short messages to indicate the availability of CP and the available data formats. The wireless bandwidth for CP is originally allocated to ego agents, and to be further assigned to co-agents when requesting collaborations. 
By jointly considering the perception topology and the wireless channel conditions, the ego agent selects an optimal subset of its nearby co-agents to share the perception data.
Compared with the simplest broadcast scheme, the proposed pull-based framework takes into account the perception need of the receiver, thereby significantly improving the communication efficiency.

% We use \emph{perception topology} to characterize the perception state of an IUS, indicating whether the transmitted data from an agent (or a group of agents) can monitor an area or detect an object.

At a co-agent, the raw sensor data needs to be processed and compressed to lightweight features to fit in the bandwidth constraint.
We adopt a two-stage detection paradigm in bird's-eye-view (BEV) representation, where perception data are projected into a global feature space with the same coordinates and semantic representations.
In this paradigm, BEV feature backbones, such as PointPillars~\cite{PointPillars} for LiDARs and Lift-Splat-Shoot (LSS)~\cite{philion2020lift} for cameras, extract RFMs from the raw sensor data. The RFM generation process serves as an effective source compression approach, extracting critical features of the interested foreground objects, while eliminating the irrelevant background information. A co-agent can flexibly adjust the data rate according to the wireless channel state, by sharing the most critical RFMs.

At the ego agent, the messages containing RFMs from multiple co-agents are decoded.
The RFMs are first spatially transformed to the local BEV coordinate of the ego agent based on the attached geo-location information.
Then, RFMs of the ego and co-agents are fused to obtain the final perception results.
Moreover, with a \emph{fuse-and-forward} mechanism, the fused BEV feature can be further shared with other agents as RFMs, implicitly facilitating a multi-hop fusion system.

Overall, the complete procedure of CP via wireless network is described in three steps:
\begin{enumerate}
	\item (Request) Based on locally available information, an ego agent sends CP requests to scheduled neighboring co-agents and specifies the maximum data rate according to the available bandwidths and measured channel states.
	\item (Response) Upon receiving a CP request, a co-agent compresses its sensor data through the channel-adaptive feature extraction algorithm according to the rate constraint, and responses with the most critical RFMs. 
	\item (Fusion)The ego agent collects and aggregates the RFMs with feature fusion. Finally, the perception result is obtained through object detection.
\end{enumerate}

To guarantee the timeliness of the perception result, the end-to-end latency is restricted to tens to hundreds of milliseconds, depending on the downstream application. Since the computation takes tens of milliseconds, communication should be finished within the remaining delay budget.

There are several performance metrics to evaluate the quality of CP.
In object detection, recall is defined as the ratio of true positive detections to all ground-truth objects, while precision is the ratio of true positive detections to all detections.
The recall and precision metrics form a trade-off, and Average Precision (AP) is the average of precision values at certain levels of recall.
Considering the contextual importance, one can also calculate a perception loss which sums up the importance-weighted penalties of false positives and false negatives.

\section{Key Challenges}

There are several key challenges in designing a task-oriented wireless communication framework for CP in IUSs, described as follows:

\subsubsection{Limited Bandwidth and Fading Channels} 
Typically, the constrained bandwidth cannot afford feature sharing among all agents in IUSs.
For example, at the sub-6 GHz frequency band, the officially allocated bandwidth for sidelink vehicular communication is only 20 MHz in China and 30 MHz in the US.
This bandwidth is shared by all vehicles for various real-time collaboration services.
Therefore, agents should extract the most valuable semantic features from the raw sensor data to share.
Besides, since the wireless channel conditions are dynamic, the achievable data rate is also time-varying.
The CP algorithm should actively adjust the data volume given the channel conditions.

\subsubsection{Inaccurate Spatio-Temporal Alignment}
Due to the multi-path effect and signal interference, localization error in IUSs is inevitable.
Moreover, the perceptual data aggregated from various perspectives may not correspond to the same timestamp, owing to stochastic sensing and communication delays.
When the ego agent extracts RFMs from its sensors and receives RFMs from others, feature-level fusion is carried out to merge those features from different perspectives. 
The common solution is to use the geo-positions and the poses of sensors, which suffers performance degradation when agents are moving and localization is inaccurate. 
Deep learning-based strategies have been proposed to compensate for the misalignment, but the lack of interpretability affects their reliability.

\subsubsection{Dynamic Perception Topology \& Channel Conditions} 
In most IUSs, the perspectives of sensors mounted on agents and the wireless communication channels are ever-changing due to the mobility of agents.
Knowledge about the perception topology and the channel conditions can stale quickly and become inaccurate.
For example, a co-agent may no longer be able to detect an interested object due to occlusion, or it may be suffering from a poor channel state and the transmission is interrupted.
If out-of-date information is used to schedule collaborators, the reliability of CP can be impaired.
Under such circumstances, the agent should quickly identify the changes in the channel conditions and the perception topology, and proceed to schedule other collaborators.
Moreover, the multi-hop mechanism may also help to aggregate information from more distant agents.

\subsubsection{Causality Issue} 
One of the most distinctive characteristics in agent scheduling is the causality issue.
The perception topology is difficult to predict because the environment is usually partially known, along with imperfect modeling of practical issues such as occlusions, sensor characteristics, and lighting conditions.
Moreover, the transmitted intermediate features from detection neural networks have the \emph{black-box nature}, which are hard to interpret without the actual execution of the networks. 
In particular, an interested object might not be correctly detected even when it is within the coverage area of a sensor, due to unusual object shapes, partial occlusions and other influencing factors.

\section{RFM Proposal and Feature Fusion}

\begin{figure*}
	\centering
	\includegraphics[width=0.75\linewidth]{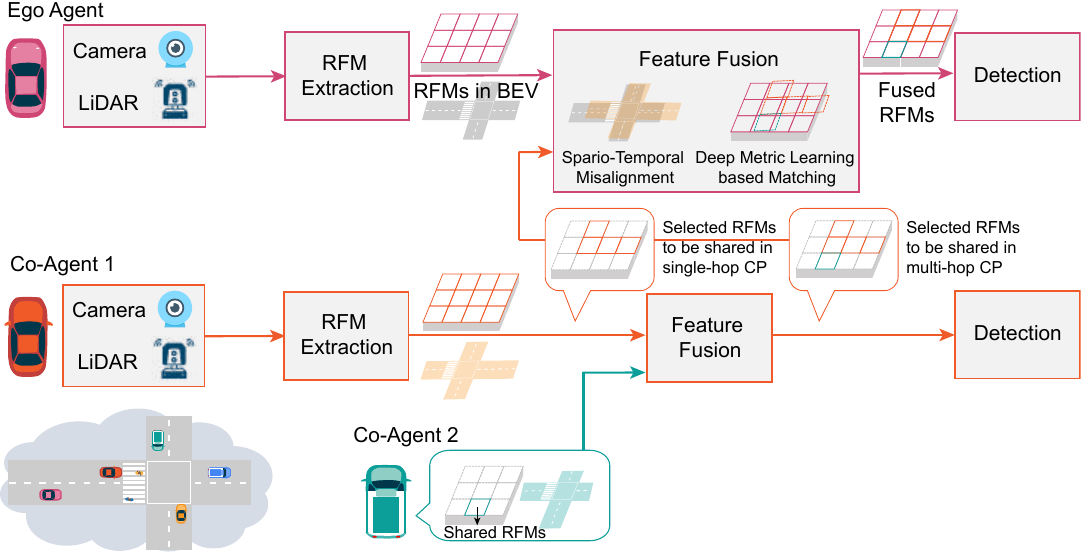}
	\caption{Illustration of the proposed channel-adaptive RFM proposal and robust feature fusion modules. }
	\label{fig:cpende}
\end{figure*}

In this section, we design feature extraction and fusion modules to overcome non-ideal channel and alignment conditions, aiming to maximize the CP gain given collaborative agents. The limited bandwidth necessitates a channel-adaptive compression scheme to extract features from raw sensor data. Meanwhile, the mobility of agents results in varying channel conditions and intermittent connectivity, which necessitates multi-hop fusion. Moreover, inaccurate spatio-temporal alignment may introduce noise into the fusion process, thus demanding robust fusion techniques to enhance calibration accuracy. The overall architecture is shown in Fig. \ref{fig:cpende}.

\subsection{Channel-Adaptive RFMs Proposal}

Through typical BEV feature backbones, such as PointPillars~\cite{PointPillars} for LiDARs and LSS~\cite{philion2020lift} for cameras, raw data are encoded and projected into a global BEV feature representation space. It provides universal coordinates among all agents, which are fundamental for multi-hop fusion. Then, the global BEV features are sent into the region proposal process to generate RFMs, along with the confidence score of each region. Noted that each RFM indicates a possible foreground area in the BEV map, the RFMs are all we need to be transmitted and fused, which greatly reduces the data volume.

To meet various communication conditions, the RFM extraction and transmission procedure should be channel-adaptive. If the channel state information (CSI) is perfect, there are some effective approaches to further compress the source data by adjusting the number of RFMs to be transmitted according to the available data rate (e.g.,~\cite{where2comm, UMC}). First, if a co-agent knows which areas have already been well-perceived by the ego agent according to the shared confidence score, it can selectively transmit the RFMs of the areas that require enhancement. Second, if agents can utilize their relative positions and historical data to infer the overlapping relationship of viewpoints, they can collaborate to perform optimized scheduling transmissions, achieving good coverage of the entire area with minimal communication cost. On the other hand, if the CSI is unknown or imperfect, the transmission procedure should take redundancy into account. The retransmission procedure as well as an error-correcting coding scheme can be utilized under such circumstances, as a future extension to the proposed source compression technique.

% For instance in Fig.~\ref{fig:Feature-level-fusion}, the wireless channel between aux vehicle 1 ({\it orange}) and the ego vehicle ({\it blue}) is in deep fading. Aux vehicle 2 ({\it green}) can first fuse the RFMs of aux vehicle 1 with its own in the global BEV representation. The feature size of the fused BEV feature map is set to be consistent with the single-view BEV feature map to ensure the consistency of subsequent multi-hop transmission. The aux vehicle 2 can then transmit the newly fused BEV feature map to the ego vehicle, which contains richer information than its own perspective. 

\subsection{Feature Fusion under Spatio-Temporal Misalignment}

To tackle the inaccurate localization (spatial) and communication latency (temporal) in the dynamic environment, we propose a robust feature map alignment strategy exploiting both semantic feature similarities and intrinsic spatial structures. The deep metric learning method is introduced to learn the similarity between different RFMs. Meanwhile, the relative positions of the RFMs from a single perspective are well-recorded under the BEV representation. Such kind of information is similar to the relative poses used in~\cite{Poses} to calibrate the localization and pose errors. Therefore, the ego agent learns to estimate the relative translation and rotation relationship between two sets of BEV RFMs by matching feature maps and spatial structures. After alignment, the agent could gain a wider perspective field from the enhanced feature maps and get more precise perception results.

\subsection{Case Study: Two-hop Feature Fusion in Connected Autonomous Driving}

We evaluate our proposed feature-level CP architecture by object detection tasks on point clouds in vehicular CP.
We use the DOLPHINS dataset \cite{Dolphins}, which contains six typical autonomous driving scenarios, such as intersections, high-way on-ramp merging, and mountain roads. For each scenario, cameras and LiDARs are equipped on three critical viewpoints, including both vehicles and road-side units.

We adopt the PointPillars~\cite{PointPillars} backbone as the global BEV feature generator. Each pillar is set to $0.16 m \times 0.16 m \times 4 m$, while the region of interest of each agent is set to be $100 m \times 100m$. Therefore, the dimension of the BEV feature maps is $625 \times 625 \times 64$. We perform a region proposal directly on the BEV feature maps, generating 300 regions that potentially contain foreground objects. Based on the foreground confidence scores, the Top-$K$ regions are selected according to current bandwidth limitations. Additionally, positioning errors following Gaussian distribution $\mathcal{N}(0,0.1^2)$ are considered in the feature fusion stage.

\begin{figure}
	\centering
	\includegraphics[width=0.9\linewidth]{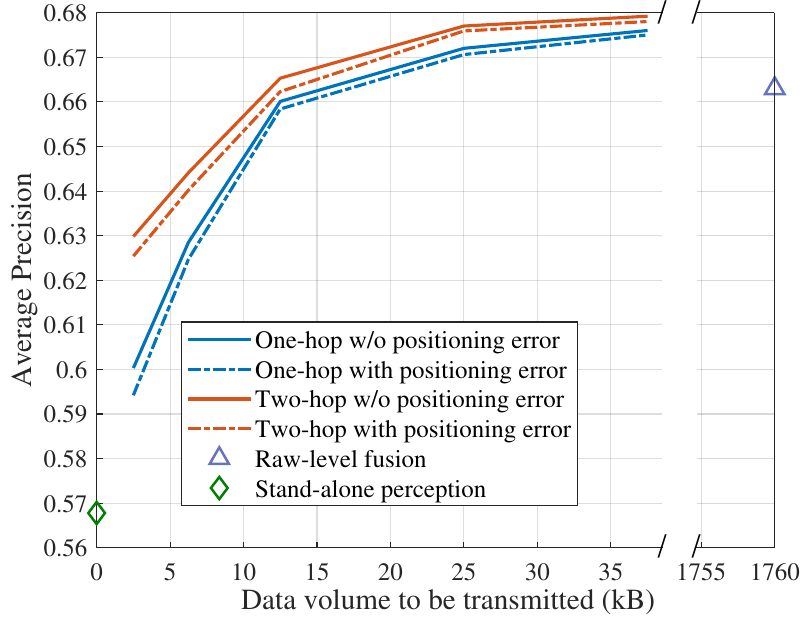}
	\caption{Performance of the proposed CP architecture.}
	\label{fig:CP_result}
\end{figure}

The experiment result is shown in Fig.~\ref{fig:CP_result}. Compared with the stand-alone perception, CP brings significant improvement under any communication payload. The gain of CP rises with an increasing number of transmitted regions, but the diminishing marginal utility reveals the possibility of saving the data volume with little precision loss. The feature-level fusion scheme also beats the raw-level fusion, since the feature extraction and region proposal stages eliminate the background noises in the raw data. On the other hand, our proposed architecture also shows robustness against positioning error, especially under large data volumes when there are more regions to be matched and aligned.

Given that the DOLPHINS dataset is recorded at 2 frames per second, a deliberate one-frame shift in the initial hop fusion introduces a delay of 0.5 seconds. Fig.~\ref{fig:CP_result} demonstrates that the collaborative fusion from a third viewpoint significantly enhances CP performance by expanding the range available to the ego agent. However, as the number of RFMs transmitted increases in the second hop, the incremental gains provided by the third viewpoint diminish. This finding underscores the critical balance between data volume and viewpoint diversity subject to limited communication bandwidth, which arouses the necessity of designing scheduling algorithms.

\section{Task-Oriented Distributed Scheduling}

Based on the RFM extraction and robust fusion modules proposed above, we further introduce task-oriented distributed scheduling algorithms in this section. The goal is to optimize the \emph{collaboration topology}, i.e., from which agents to request RFMs.

To make scheduling decisions, it is fundamental to define the \emph{collaboration utility} associated with an agent or a group of agents.
Depending on the metric of CP, this utility could be the extra coverage area or additionally detected objects.
The perception topology is the crucial factor to the utility, parameterized by the transmission bandwidth, since more data generally assist the detection of objects, as reported in the previous section.
Therefore, in bandwidth-constrained networked IUSs, scheduling more agents can sometimes negatively influence the performance of CP, because the transmitted data from each agent becomes less comprehensive.
This intuition necessitates a task-oriented wireless communication system with a \emph{sparse and essential} collaboration topology to optimize perception performance.

To address the design challenges, we review existing solutions in the literature and propose an efficient yet effective approach, learning while scheduling.

\subsection{Spatial Reasoning}
To obtain an approximated perception topology, the fundamental approach is spatial reasoning based on geometry relationships \cite{Autocast}.
Specifically, based on the locally available sensor data, the agent can construct a dynamic environment map and calculate the visibility of an object from another nearby agent.
The visibility is determined by the line-of-sight condition, indicating whether there are blockages between the object and the agent.
This spatial reasoning process can be enhanced with the assistance of a high-definition map, which provides the background information of the environment.

Although efficient to implement, spatial reasoning inevitably introduces error since there can be blockages that are unobserved by the agent.
As stated previously, visibility alone is insufficient for accurate prediction of real-world object detectors, which results in unexpected false negatives.
Furthermore, when multi-hop feature fusion is introduced, the collaboration utility of an agent is beyond its visibility.
Therefore, spatial reasoning has its limitations in evaluating modern CP systems. 
% especially when dealing with feature-level sensor data.

\subsection{Metadata Exchange}
Inspired by the attention mechanism, a three-stage handshake communication is designed to determine the utility of CP \cite{when2comm}.
The perceptual data of each agent is first compressed into compact query and key features, then the queries are broadcast to all neighboring agents to compute the query-key matching scores.
Finally, the priority of collaboration can be decided by descending order of the scores.
Note that the feature extraction and matching processes are functions with learnable parameters, which generally capture the relevance between the sensor data.
% Subsequently, this mechanism restricts its application to augmentation of degraded but visible observations because the matching process cannot distinguish between invisibility and irrelevance.

In \cite{where2comm}, a BEV spatial confidence map is first encoded from the sensor data, which reflects the probability that there is an object within the area.
In the first receiver-agnostic collaboration round, a small portion of features is broadcast based on the confidence scores, along with a request map that negatively correlates with the confidence map.
For the following rounds, the utility of transmitting an area is calculated by the multiplication of the sender's confidence and the receiver's request map, which represents the potential of discovering missed information.
This approach lays the foundation for fine-grained sensor data distributed scheduling at the instance level.
However, the above solutions based on metadata exchange introduce extra communication and computation overheads that result in higher end-to-end latency, which in turn impairs the perception.

\subsection{Learning While Scheduling}
Instead of requesting extra information, we propose to leverage the historical data from nearby agents to learn the CP utilities at present.
In a task-oriented way, the utility of received features from an agent can be extracted from the innovative corrections of CP, by comparing the perception results of CP with stand-alone perception.
The tricky part is that the ego agent can only evaluate the utilities after it completes the CP in this time frame, while the potential utilities of unscheduled agents are evolving and unobservable.
With the utility knowledge of previous feature maps, the agent can proactively schedule co-agents to continuously \textit{exploit} the sensor data, and update the knowledge at the same time.
Note that with the multi-hop mechanism, the feature map of a co-agent could also include the sensing information from its neighbor agents.
In a highly dynamic scenario, one has to \textit{explore} by scheduling each candidate co-agent once in a while to learn the perception topology, otherwise one could miss a co-agent that provides potentially superior sensor data.
This forms an exploration-exploitation trade-off.

\begin{figure}
	\centering
	\includegraphics[width=\linewidth]{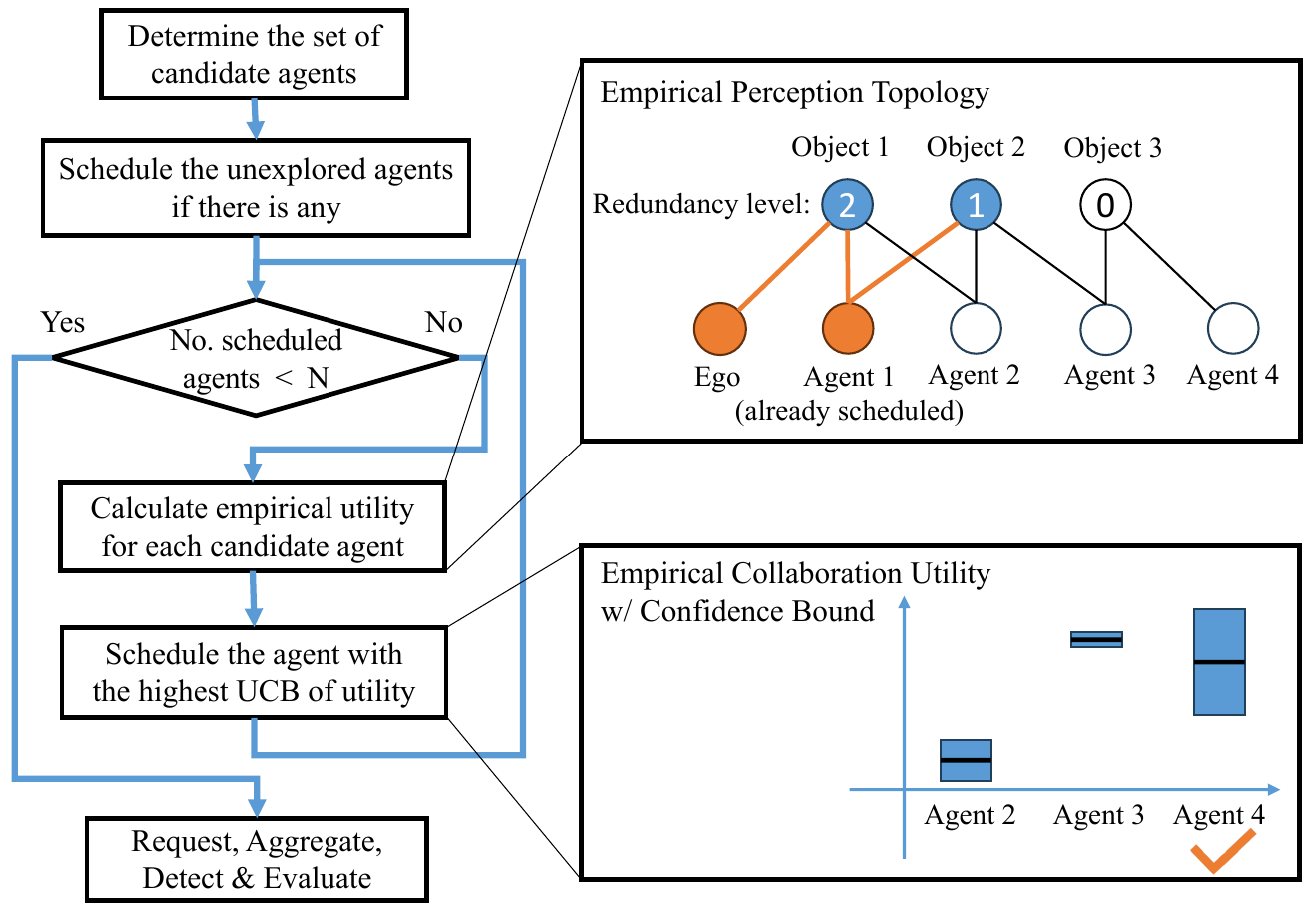}
	\caption{Illustration of the C-MASS scheme. 
     \emph{Left}: The diagram of C-MASS algorithm. 
     \emph{Upper right}: The orange nodes and edges are already scheduled agents and anticipated detections in the current round. Empirical collaboration utilities are calculated using the learned perception topology and the current redundancy level.  
     \emph{Bottom right}: The confidence bounds of collaboration utilities. Agent 4 is preferred over Agent 3 because it has not been explored for a relatively long time.}
	\label{fig:c-mass}
\end{figure}

In a simpler setting that considers scheduling only one collaborator at a time, our preliminary work adopts the theoretical framework of restless multi-armed bandits (RMAB) \cite{MASS-TVT}.
One of the most straightforward solutions is the periodic explore-then-commit (ETC) algorithm. 
In ETC, each nearby agent is scheduled once at the beginning of an epoch, and the empirical leader is scheduled constantly until the end of the epoch.
However, it is unable to adapt to the rapid and irregular shifts of the utilities.
Inspired by the Upper Confidence Bound (UCB) algorithm in the static MAB problem, the mobility-aware sensor scheduling (MASS) algorithm is proposed to balance the exploration-exploitation trade-off in restless bandits problems.
To tackle the dynamics of utilities, Brownian motion is used to approximate the underlying process of the utilities.
The UCBs of collaboration utilities, swelling through time when unexplored, are ranked to encourage moderate exploration.
If a new co-agent is discovered, it is scheduled immediately, and its UCB of utility will be computed in subsequent scheduling decisions until it leaves.

Extending to scheduling multiple collaborators, a combinatorial mobility-aware sensor scheduling scheme, C-MASS, is proposed.
To reduce the complexity due to the combinatorial explosion, we exploit the problem structure and adopt a multi-round greedy approach, as illustrated in Fig. \ref{fig:c-mass}.
In each round, the ego agent calculates the empirical collaboration utilities of the candidates based on the learned perception topology from past observations.
In specific, the utility consists of new detections from the candidate alone and also from the collaborative detections with the other agents.
We also note that new anticipated detections contribute to the empirical utility according to other factors such as the importance and the current redundancy level of the object.
Taking the dynamics of the environment into account, we then compare the UCBs of CP utilities to greedily schedule the most beneficial agent to collaborate in this round.
After $N$ rounds of selection, the ego agent requests sensor data and executes the detection network.
Finally, the perception topology of the scheduled agents is evaluated and updated for future scheduling.
In addition to the task-oriented characteristic, the proposed C-MASS scheme also has the merit that no extra communication overhead is required prior to scheduling decisions.

\subsection{Case Study: Task-oriented Collaborative Vehicle Selection in Connected Autonomous Driving}

To evaluate and compare the solutions, we conduct a case study in a connected autonomous driving scenario where collaborative vehicles (CoVs) exchange sensor data to achieve a more holistic environmental perception.
The CoVs periodically beacon short status messages which also declare their functionality of CP.
Under a limited wireless bandwidth, the ego vehicle selects a subset of nearby available CoVs, requests and aggregates their sensor data, and finally executes the detection network, at an operating frequency of 10Hz.
Each undetected traffic participant incurs a perception loss that equals its importance weight, which relates to the distance to the ego vehicle.
Therefore, the utility of CP is defined as the sum of weights of the additionally detected objects.
In this case study, we consider maximizing the average utility of CP by scheduling $N$ CoVs in a time frame and allocating equal bandwidths to the CoVs.

For simulation, we set up an urban traffic setting similar to \cite{MASS-TVT}, where vehicles and pedestrians are moving along bi-directional two-lane streets with sidewalks.
Assume the market penetration ratio (MPR) is $0.7$, which denotes the ratio of CP-enabled CoVs among all the vehicles.
We adopt the 3GPP V2V sidelink channel model specified in 3GPP TR 37.885 \cite{3gpp37885}. The link is classified by LOS, NLOSv, and NLOS, with different sets of parameters. The channel gain of a V2V link consists of a distance-related path loss, a spatial-correlated shadowing loss, and attenuation losses caused by vehicle blockages. The fast fading is simulated with i.i.d. Rician distribution for LOS and NLOSv links, and Rayleigh distribution for NLOS links.
Under the bandwidth constraint $0.6\text{MHz}$, the proposed C-MASS scheme is compared with two baselines:
\textbf{Closest Candidates} scheme selects the closest $N$ CoVs, which are generally close to the most important objects and enjoy the best communication channels.
\textbf{Greedy Coverage} greedily maximizes the weighted coverage area in each of the $N$ rounds. 
However, accurate calculation of coverage requires high-definition maps and spatial reasoning which induce extra delay in scheduling decisions.
When the number $N$ is large, all three schemes converge to \textbf{All Candidates}, which selects all candidate CoVs within the communication range of $150\text{m}$.

\begin{figure}[t]
	\centering
	\subfloat[]{\includegraphics[width=0.9\linewidth]{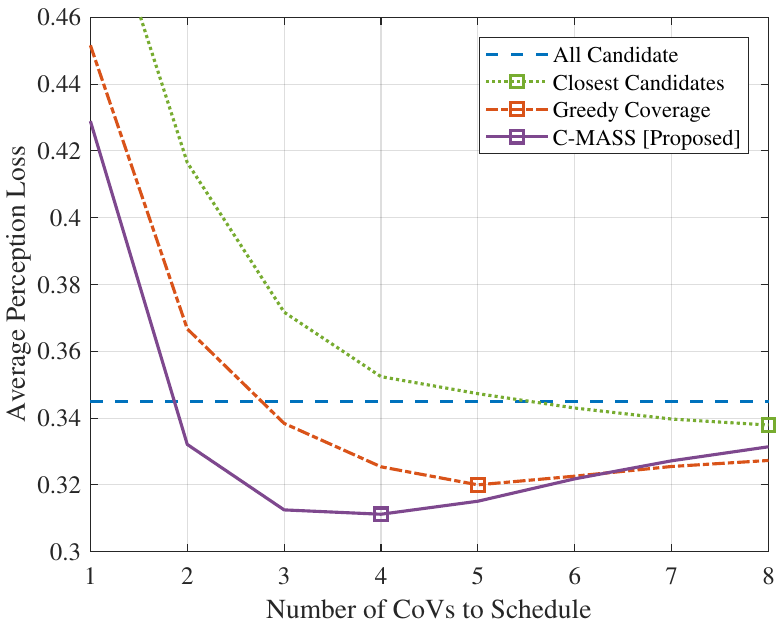}
		\label{subfig:perception-loss}}
	\hfill
	\subfloat[]{\includegraphics[width=0.9\linewidth]{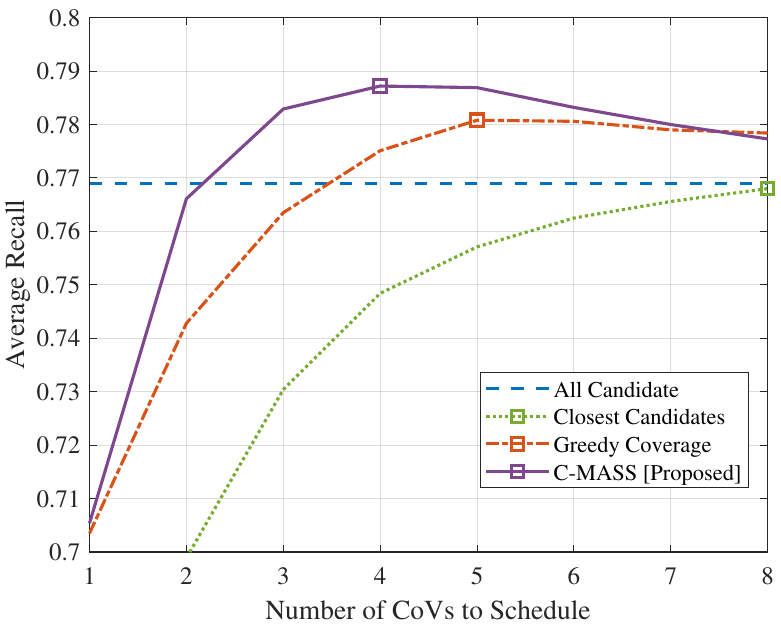}
		\label{subfig:recall}}
	\caption{Performance of CP with different CoV selection schemes. (a) The average perception loss. (b) The average recall.}
	\label{fig:perf-sel}
\end{figure}

The experiment result with a varying number $N$ of selected CoVs is shown in Fig. \ref{fig:perf-sel} during a trip of 10,000 frames (1,000 seconds).
As expected, with larger $N$, the bandwidth allocated to each CoV is less, which in turn negatively influences the utility of CP.
It can be seen from Fig.~\ref{fig:perf-sel} that the optimal number of selected vehicles are $N=4$ for C-MASS, $N=5$ for Greedy Coverage, and $N=8$ for Closest Candidates, respectively.
Moreover, C-MASS outperforms all the baselines in terms of both the perception loss and the recall metrics.
Intuitively, C-MASS is oriented directly towards detecting more important objects rather than keeping wider coverage.
Although the empirical perception topology is from past observations, UCB is introduced to effectively balance exploration and exploitation in a dynamic environment, as discussed in \cite{MASS-TVT}.
The proposed scheme is also highly efficient without the need to acquire metadata such as coverage information.

\section{Conclusion and Outlook}

In this article, we have proposed a task-oriented wireless communication framework for CP, addressing the challenges of limited bandwidth, inaccurate spatio-temporal alignment, dynamic topologies, and the causality issue. The framework consists of BEV-based RFM extraction, metric learning-based fusion, and task-oriented distributed agent scheduling. Case studies showcase the perception performance gain brought by the proposed framework, and also inspire some future directions as follows.

\textbf{CP Timeliness.} The study of robust feature fusion highlights the introduction of communication latency during multi-hop feature fusion. As the first case study shows, the more hops a set of RFMs needs to be transmitted, the less informative it is, and thus the confidence score is decayed according to the freshness. One possible solution is to use metrics like age of information to represent the time decay of each fusion process. In addition, the computation delay should also be appropriately modeled and considered.   

\textbf{Energy Efficiency Optimization.} The end-to-end process of CP not only consumes the wireless communication energy but also the computing energy. While one can save communication energy by sharing less data, the energy consumption for feature extraction and compression, as well as for fusion may increase. Careful design should be adopted to balance the energy consumption from communications and computing under given timeliness constraints, to achieve better energy efficiency. 

\textbf{Security Issues.} CP over wireless inevitably needs to share sensor information among agents, which leads to potential privacy leakage. In-depth analysis on the privacy information carried by RFMs is required, based on which coding schemes and protocols should be designed for better privacy preserving. Since malicious agents may send fake sensor data or RFMs, consensus algorithms among agents are needed to detect such harmful behaviors.

\end{document}